\documentclass[12pt]{iopart}
\usepackage{iopams}  
\usepackage{epsfig}  
\usepackage{pstricks}
\usepackage{rotate}
\usepackage{psfig}

\begin{document}

\newcommand{\gapprox}{\stackrel{>}{_{\sim}}}
\newcommand{\lapprox}{\stackrel{<}{_{\sim}}}
\newcommand{\ra}{\rightarrow}
\newcommand{\ccb}{c\bar{c}}
\newcommand{\bbb}{b\bar{b}}
\newcommand{\pbp}{\bar{p}p}
\newcommand{\epem}{e^+e^-}
\newcommand{\gaga}{\gamma\gamma}
\newcommand{\ftc}{F_2^{c}}
\newcommand{\ptr}{p_T^{rel}}
\newcommand{\pbmo}{pb$^{-1}$}
\newcommand{\rk}{\mbox{\boldmath $k$}}  
\newcommand{\rkn}{\mbox{$k$}}  
\newcommand{\rr}{\mbox{\boldmath $r$}} 
\newcommand{\rp}{\mbox{\boldmath $p$}} 
\newcommand{\rqq}{\mbox{\boldmath $q$}} 
\title[Heavy Quark Photoproduction in $\rk_{\perp}$-Factorization Approach]
      {Heavy Quark Photoproduction in $\rk_{\perp}$-Factorization Approach}

\author{C. Brenner Mariotto, M.B. Gay Ducati, M.V.T. Machado}

\address{ Instituto de F\'{\i}sica, Universidade Federal do Rio Grande do
Sul. \\  Caixa Postal 15051, 91501-970 Porto Alegre, RS, BRAZIL.}

\begin{abstract}
We investigate  the heavy quark photoproduction based on the $\rk_{\perp}$-factorization approach, focusing on the results from the saturation model. The deviations in the results using the unintegrated gluon distribution considering the saturation model and the derivative of the collinear gluon distribution are analysed. Total cross sections and $p_T$ distributions are analysed in detail, setting the deviations between the color dipole approximation and the complete semihard approach. 
\end{abstract}

\section{Introduction}

The investigation of heavy-quark production in high energies provides 
a better understanding of the hadron internal structure. In particular, the heavy quark masses are large enough to be taken as a hard scale, making the strong coupling constant small and allowing a perturbative QCD treatment of the process.
Heavy quarks are produced in the clean $\gamma p \rightarrow Q\bar{Q}$ reaction 
at HERA, where the incident photon (real or virtual) probes the proton target at high center of mass energy $W$. There is a large amount of data 
on heavy flavor production at HERA, which have been a plethora on
studies analyzing the interface between hard and soft regimes \cite{review1,review1b}. However, data on
open heavy quark production are limited by small statistics and the
intermediate energy interval between the fixed target experiments and the high
energy at HERA have not been covered yet. In hadroproduction at the $p\bar{p}$ Tevatron collider, 
the situation is slightly better, having precise measurements of the transverse momentum distribution of the produced heavy quarks \cite{review2,review3}.  
 
The underlying  mechanism for heavy quark production at HERA 
is photon gluon fusion: a photon coupled to the scattered electron
interacts with a gluon from the proton by producing a quark antiquark 
pair, e.g. $\ccb$ (charm) or $\bbb$ (bottom). In the collinear QCD approach, based on the well known collinear factorization theorem \cite{collfact}, the process is described through the convolution of on-shell matrix elements, encoding the partonic subprocesses, with the  parton distribution functions. At high energies, 
the gluon is the parton which drives the dynamics, 
evoluting on the virtualities obeying the DGLAP evolution equations \cite{DGLAP}. 
The transverse momenta of the incident particles are taken as zero and in the computation of the cross sections one averages over two transverse polarizations of the initial gluon. 
A computation of the process requires the knowledge of the gluon momentum distribution in the proton and the calculation of the $\gamma g$ subprocess. The gluon density as a function of the longitudinal momentum fraction  $x$ and  virtuality scale $Q^2$ is known to an accuracy of a few percent from a global analysis of 
scaling violations of the proton structure function $F_2$ 
measured at HERA \cite{CTEQ5,MRST,GRV98}. 
The QCD matrix elements of the partonic process have been calculated 
in next-to-leading order (NLO) accuracy \cite{fmnr}. Still substantial theoretical uncertainties come from the 
heavy quark mass and the from 
renormalization and factorization scales, $\mu_R$ and $\mu_F$. 

The collinear factorization approach has produced a successful description of 
single-particle distributions and total cross sections for heavy-quark production. On the other hand, despite of many phenomenological successes, results within this approach are in contradiction with data on azimuthal correlations and on distributions in transverse momentum of the produced heavy-quark pair \cite{RSS} (for quarkonium production see {\it e.g.} Refs. \cite{review1b,MGID}). This problem is in general cured by introducing an intrinsic transverse momenta distribution (intrinsic $k_\perp$) for the incident partons, parametrized using  a gaussian profile. However, the mean value of intrinsic $k_\perp$ needed to describe azimuthal correlations and $p_T$ spectrum can be as high as 
$\langle \rkn_{\perp}\rangle \simeq 1$ GeV or even $2 - 3$ GeV, which is not suitable from non-perturbative arguments. 
Moreover, recent calculations using known NLO results \cite{NLOppmassive} of bottom hadroproduction in the collinear approach undershoot Tevatron data \cite{bcdfd0} by a factor $2$ or $3$, suggesting that an important contribution to the computed observables is missing.

At high energies another  factorization theorem emerges, the $\rk_{\perp}$-factorization or semihard approach \cite{CCH,CE,GLRSS}. The relevant diagrams are considered with the virtualities and polarizations of the initial partons, taking into account the transverse momenta $q_{1\perp}$ and $q_{2\perp}$ of the incident partons. 
The processes are described through the convolution of off-shell matrix elements with the unintegrated parton distribution, ${\cal F}(x,\rk_{\perp})$. The latter can recover the usual parton distributions in double logarithmic limit (DLL) by its integration over the transverse momentum $\rk_{\perp}$. In the asymptotic energy  limit, the unintegrated gluon distribution should obey the BFKL evolution equation \cite{BFKL}. 
At the present, there is a lack of an accurate determination of this quantity at the same level as the usual parton densities. The matrix elements computed for the relevant subprocesses within the $\rk_{\perp}$-factorization approach are more involved than those needed in the collinear approach already at LO level. On the other hand, a significant part of the NLO and some of the NNLO corrections to the LO contributions on the collinear approach, related to  the  contribution of non-zero transverse momenta of the incident partons, are already included in the LO contribution within the $\rk_{\perp}$-factorization formalism \cite{RSS}. Moreover, part of the virtual corrections can be  resummed in the unintegrated gluon function \cite{RSS}. Recently, the computation of NLO corrections to the subprocesses have been completed and full calculations at this level should appear in the near future \cite{bartels-nlo}. An important feature of the approach is the equivalence at 
leading logarithmic approximation with the color dipole picture \cite{Bialas}, which have been used in a profuse phenomenology at HERA \cite{McDermott}. Besides, a very important issue is the consistency of the $\rk_{\perp}$-factorization approach including nonleading-log effects with the collinear factorization beyond leading order \cite{catani-hautmann}: 
the coeficient functions and the splitting functions providing $q(x,Q^2)$ and $G(x,Q^2)$ are supplemented with the all-order resummation of the $\alpha_s\ln (1/x)$ contributions at high energies, in contrast with a calculation in fixed order perturbation theory.

Two additional ingredients should be taken into account when considering the semihard approach: the infrared sector and saturation effects. The unintegrated gluon function should evolve in tranverse momentum through the BFKL evolution at high energies, leading to the diffusion on $\rk_{\perp}$ of the initial gluons in the evolution process \cite{Forshaw:1997wn}. In this diffusion scenario 
the transverse momenta values are spread out into the infrared (and ultraviolet) region, where the perturbative description is not 
completely reliable. Therefore, the evolution should take into account properly the correct behavior in that region. 
The recently calculated non-linear corrections to the BFKL approach \cite{GB-Motyka-Stasto} introduce a natural treatment for these difficulties, where the saturation scale $Q_s$ provides the suitable cut-off controlling the infrared problems. As the longitudinal momentum fraction $x\simeq Q^2/W^2$ decreases, unitarity corrections become important and 
control the steep growth of the gluon distribution. In this domain the gluon distribution can saturate completely or acquire a mild logarithmic increasing. The most appealing approach taking into account both the notions of infrared behavior (confinement) and parton saturation phenomenon  is the saturation model \cite{GBW}, which is an eikonal-type model based on the color dipole picture of high energy interactions. In this approach the physical picture is quite simplified and the expression for the saturation scale is promptly calculated. The adjustable parameters of the model are obtained from a fit 
to small-$x$ HERA data on the inclusive structure function and the photoproduction total cross section, being suitable for further applications to more exclusive quantities as open heavy-quark production.

This paper is organized as follows. In the next section we present the main formulae for the heavy-quark photoproduction within the  $\rk_{\perp}$-factorization approach, defining the relevant kinematical variables. Furthermore, we investigate the unintegrated gluon distribution 
from the saturation model in comparison with the results from the derivative of the integrated 
gluon distribution, $x{\cal G}(x,\rk_{\perp})$. In Sec. (\ref{sect:3}), we present our results on the total cross section for charm and bottom photoproduction, as well as estimates for the $p_{\perp}$-distribution of the produced heavy quark pair.
We investigate in detail the deviations in the results when one confronts the dipole approximation and a conservative $\rk_{\perp}$-factorization procedure. Namely, the choice of scale for the coupling constant and the suitable longitudinal momentum fraction entering in the unintegrated gluon function. The predictions from the derivative of the collinear gluon function are also studied. In the last section, one presents the conclusions and final considerations.

\section{Heavy-quark photoproduction in the $\rk_{\perp}$-factorization}
\label{sect:2}
In this section we investigate the quasi-real photon scattering off a proton in the semihard approach. In light-quark photoproduction there is a poor 
knowledge concerning the effective light quark mass, which is associated with non-perturbative aspects of the process. This problem is naturally solved in 
heavy-quark photoproduction, because of the heavy quark mass. 
The semihard approach is valid in the domain where the following double inequality holds: $s\gg \mu^2 \simeq \hat{s} \gg \Lambda_{QCD}^2$, i.e. the typical parton interaction scale $\mu^2$ is much higher than the QCD cutoff parameter $\Lambda_{QCD}^2$, and much lower than the center of mass energy, $\sqrt{s}$. The $\rk_{\perp}$-factorization approach can resum in the leading logarithmic approximation all the contributions proportional to $[\alpha_s\ln (\mu^2/\Lambda_{QCD}^2)]^n$, $[\alpha_s\ln (\mu^2/\Lambda_{QCD}^2)\ln (1/x)]^n$ and $[\alpha_s\ln (1/x)]^n$, where the first one  corresponds to the collinear DGLAP resummation \cite{DGLAP}, the second one the double logarithmic contribution and the last one the BFKL resummation \cite{BFKL}. Such resummation leads to the unintegrated gluon distribution, ${\cal  F}(x, \rk_{\perp})$, which can also depend on the scale $\mu^2$. It gives the probability to find a parton carrying a longitudinal momentum fraction $x$ and transverse momentum $\rk_{\perp}$. Requiring that in the DLL limit, the transverse photoabsortion total cross section written in the   $\rk_{\perp}$-factorization approach to be consistent with the same limit from the DGLAP approach, the collinear gluon distribution is given by the unintegrated one in the following way,
\begin{eqnarray}
x\,G(x,\mu^2)= \int^{\mu^2} \frac{d\rkn_{\perp}^2}{\rkn_{\perp}^2}\, {\cal F}(x,\rkn_{\perp}^2)\,,
\label{eq:1}
\end{eqnarray}
where $\mu^2$ is the scale of the process, which can be $Q^2$, the heavy-quark mass or the $p_{\perp}$ of the produced particles, for instance. In particular, in obtaining Eq. (\ref{eq:1}), a strong ordering condition $\rk_{\perp}^2\ll \rp_{\perp}^2 \ll \mu^2$ is considered, where $ \rp_{\perp}$ is the transverse momentum of the quark-antiquark loop in the photon vertex.

\begin{figure}[t]
\begin{center}
\scalebox{0.7}{\includegraphics*[198pt,503pt][540pt,753pt]{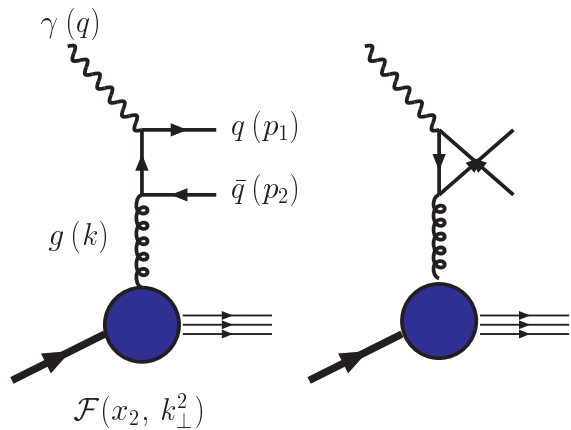}}
\caption{\label{diagrams} The leading order QCD diagrams for heavy-quark production via photon-gluon fusion. The momenta of the particles are shown and 
the blob represents the gluon emission chain encoded in the unintegrated gluon distribution, ${\cal F}(x_2,\rk_{\perp}^2)$.}
\end{center}
\end{figure}

In order to compute the cross section of a physical process \cite{LiSaZot2000,Shabel-Shuva}, the unintegrated gluon function should be convoluted with the off-shell matrix elements for the relevant partonic subprocesses. In these matrix elements the polarization tensor of the virtual gluon is given by $L_{\mu \nu}^{(g)}=\overline{\varepsilon^{\mu}_g \, \varepsilon^{*\nu}_g}=\rkn_{\perp}^{\mu} \rkn_{\perp}^{\nu}/|\rk_{\perp} |^2$. In the following we will calculate the total and differential cross section of heavy-quark photoproduction (charm and bottom) taking into account the diagrams shown in Fig. (\ref{diagrams}). For convenience, one can define the Sudakov variables for the process $ep \rightarrow Q\bar{Q}X$ at high energies,
\begin{eqnarray}
p_1 & = & \alpha_1 P_1 + \beta_1 P_2 + \rp_{1\perp}\,, \hspace{2cm} q  =  x_1 P_1 +  \rqq_{\perp}\,, \label{eq:2}\\
p_2 & = & \alpha_2 P_1 + \beta_2 P_2 + \rp_{2\perp}\,,  \hspace{2cm} \rkn  =  x_2P_2 + \rk_{\perp}\,,\label{eq:3}
\end{eqnarray}
where, as it is shown in Fig. (\ref{diagrams}), $p_1$ and $p_2$ are the four-momenta of the produced heavy-quarks, $q$ and $\rkn$ are the photon and the gluon four-momenta, respectively. 
The corresponding transverse momenta are $\rp_{1\perp}$, $\rp_{2\perp}$, $\rqq_{\perp}$ and $\rk_{\perp}$. The electron and proton momenta are denoted by $P_1$ and $P_2$. In the center of mass frame of the process, one has 
$P_1=(\sqrt{s}/2,0,0,\sqrt{s}/2)$, $P_2(\sqrt{s}/2,0,0,-\sqrt{s}/2)$, $P_1^2=P_2^2=0$ and $(P_1\cdot P_2)=s/2$, with $\sqrt{s}$ being the center of mass energy. From simple inspection of Eqs. (\ref{eq:2}) and (\ref{eq:3}) and using the energy-momentum conservation law, one obtains the relations,
\begin{eqnarray}
p_1^2 & = & p_2^2 = m_Q^2 \,, \hspace{0.6cm} q^2=\rqq_{\perp}^2\,,  \hspace{0.6cm} \rkn^2 = \rk_{\perp}^2\,, \label{eq:4}\\
 \rqq  _{\perp}& + & \rk_{\perp}  =   \rp_{1\perp} +  \rp_{2\perp}\,.\label{eq:5}
\end{eqnarray}

The Sudakov variables can be written in terms of the transverse masses $m_{1,2\,\perp}^2= m_Q^2 +  \rp_{1,2\,\perp}^2$, where $m_Q$ is the heavy-quark mass{\bf ,} and the heavy-quark rapidities $y^*_{1,2}$, in the following way
\begin{eqnarray}
\alpha_1 &  = &  \frac{m_{1\perp}}{\sqrt{s}}\,\exp(y_1^*)\,, \hspace{1cm} \alpha_2 = \frac{m_{2\perp}}{\sqrt{s}}\,\exp(y_2^*)\,, \label{eq:6}\\
\beta_1 & = &  \frac{m_{1\perp}}{\sqrt{s}}\,\exp(-y_1^*)\,, \hspace{0.7cm} \beta_2 = \frac{m_{2\perp}}{\sqrt{s}}\,\exp(-y_2^*)\,, \label{eq:7}\\
x_1 & = & \alpha_1 + \alpha_2 \,, \hspace{2.0cm} x_2= \beta_1 + \beta_2 \,.\label{eq:8}
\end{eqnarray}
where in the photoproduction case $q=P_1$, the variable $x_1$ simplifies to $x_1=1$ and one can define $\alpha_1\equiv z$ and $\alpha_2\equiv 1-z$. The $z$ and $(1-z)$ variables correspond to the longitudinal momentum fraction carried by the heavy-quark having transverse momentum $\rp_{1\perp}$ and $\rp_{2\perp}$, respectively.

Having introduced the relevant definitions and variables, the differential cross section for the photoproduction process is expressed as the convolution of the unintegrated gluon function with the off-shell matrix elements  \cite{CataCiafaHaut,CCH,LiSaZot2000},
\begin{eqnarray}
  \frac{d\sigma (\gamma p \rightarrow Q\bar{Q}X)}{d^2 \rp_{1\perp}}   = \int dy_1^* \,d^2 \rk_{\perp} \frac{{\cal F}(x_2,\rk^2)\,\, |{\cal M}|^2(\mathrm{off\!-\!shell})}{\pi \alpha_2} \,,\label{eq:9}
  \end{eqnarray}
where the off-shell LO matrix elements are given by \cite{CataCiafaHaut,LiSaZot2000},
\begin{eqnarray}
|{\cal M}|^2 (\mathrm{off\!-\!shell})  &  = &   \alpha_{em}\,\alpha_s(\mu^2)\,e_Q^2 \,  \left[   \frac{z^2 + (1-z)^2}{[(p_2-\rkn)^2 - m_Q^2]\,[(p_1-\rkn)^2-m_Q^2]} \right. \nonumber \\
  & + &  \left.  \frac{2m_Q^2}{\rkn_{\perp}^2}\,\left( \frac{z}{(p_1-\rkn)^2-m_Q^2} - \frac{1-z}{(p_2-\rkn)^2 - m_Q^2}      \right)^2\, \right] \,,\label{eq:10}
\end{eqnarray}
where $\alpha_{em}=1/137$ is the electromagnetic coupling constant and $e_Q$ is the electric charge of the produced heavy-quark. The scale $\mu$ in the strong coupling constant will be specified latter on. In general, it is taken to be equal to the gluon virtuality, $\mu^2=\rk^2$ in close connection with the BLM scheme \cite{BLM}. In the leading $\ln (1/x)$ approximation, $\alpha_s$ should take a constant value. When the transverse momenta of the incident partons are sufficiently smaller than those from the produced heavy-quarks, the result from the collinear approach is recovered. The final expression for the photoproduction total cross section considering the direct component 
of the photon can be written as \cite{Shabel-Shuva}
\begin{eqnarray}
&\sigma_{tot}^{phot}& 
= 
  \frac{\alpha_{em}\,e_Q^2}{\pi}\, \int\, dz\,\,d^2 \rp_{1\perp} \, d^2\rk_{\perp} \, \frac{\alpha_s(\mu^2)\,{\cal F}(x_2,\rk_{\perp}^2; \mu^2)}{\rk_{\perp}^4}\nonumber \\
&&\!\!\!\!\!\!\times 
 \left\{ [z^2+ (1-z)^2]\,\left( \frac{\rp_{1\perp}}{D_1} + \frac{(\rk_{\perp}-\rp_{1\perp})}{D_2} \right)^2 +   m_Q^2 \,\left(\frac{1}{D_1} + \frac{1}{D_2}  \right)^2  \right\}\,, \label{eq:11} 
\end{eqnarray}
where $D_1 \equiv \rp_{1\perp}^2 + m_Q^2$ and 
$D_2 \equiv (\rk_{\perp}-\rp _{1\perp})^2 + m_Q^2$.

In Eq. (\ref{eq:11}) the unintegrated gluon function was allowed to depend also on the scale $\mu^2$, since some parametrizations take this scale into account in the computation of that quantity (see, for instance Ref. \cite{lund-small-x} for a compilation of a number of them). We are now ready to calculate the total and differential cross sections for the process, provided a suitable input for the function ${\cal F}(x_2,\rk_{\perp}^2; \mu^2)$. The practical procedure in this paper is to consider one of the simplest parametrizations available, covering a consistent treatment of the infrared region and taking into account the expected saturation effects  at high energies. These features are fullfilled 
in the saturation model, which it is shortly reviewed in the following and it will be conveniently contrasted with the results coming from the derivative of the collinear gluon function,
\begin{eqnarray}
x \, {\cal G} (x,\rk_{\perp}^2)= \frac{\partial\, [\,x\,G(x,\rk_{\perp}^2)\,]}{\partial \, \ln \rk_{\perp}^2}\,,\label{eq:13}
\end{eqnarray}   
where $G(x,\mu ^2)$ is the integrated gluon distribution, which can be taken from the available parameterizations in the literature \cite{CTEQ5,MRST,GRV98}. 

\subsection{The saturation model}
\label{subsec:21}

The saturation model \cite{GBW} is based on the color dipole picture of the interaction, 
represented in the target rest frame where the transverse size $\rr$ of the dipole quark-antiquark coming from the virtual photon  Fock state fluctuations is fixed during the interaction. This representation can be recovered from the $\rk_{\perp}$-factorization approach in the leading logarithmic approximation through the Fourier transform between the transverse size  and the transverse momentum spaces. The photoabsortion total cross section is written as a convolution of the virtual photon wavefunction with the effective cross section for the interaction dipole-target,
\begin{eqnarray}
\!\!\!\!\!\!\!\!\!\!\sigma_{tot}^{\gamma^* p} (x,Q^2)=\int dz\, \int d^2 \rr \, \left(|\Psi_T(z,\rr,Q^2)|^2 +  |\Psi_L(z,\rr,Q^2)|^2\right) \, \sigma_{dip}(x,\rr)\,, \label{eq:14}
\end{eqnarray} 
where the dipole cross section interpolates between the color transparency behavior for small size dipole configuration and the confinement features for large dipole sizes. It reads
\begin{eqnarray}
\sigma_{dip} & = &  \sigma_0\, \left[\, 1- \exp \left( -\frac{\rr^2}{4\,R_0^2(x)} \right)\,  \right]\,, \label{eq:15}\\
R_0^2(x)& = & \frac{1}{\mathrm{GeV}^2}\, \left( \frac{x}{x_0}\right)^{\lambda}\,, \label{eq:16}
\end{eqnarray}
where $R_0(x)$ is the saturation radius which decreases when $x\rightarrow 0$. The parameters $\sigma_0$, $x_0$ and $\lambda$ are determined from a fit to small-x 
HERA data.  
An additional parameter is the effective light quark mass $m_q=140$ MeV, which is needed to produce finite results on the photoproduction total cross section. The saturation scale is defined as $Q_s^2=1/R_0^2(x)$: when $r\,Q_s/2\ll 1$ the model reproduces color transparency, $\sigma_{dip}\sim r^2$, whereas when $r\,Q_s/2\geq 1$ the cross section tends to a constant at large $r\,Q_s$ (it simulates confinement), $\sigma_{dip}\sim \sigma_0$. In the region $r\,Q_s/2\sim 1$, the model simulates the physics from the multiple scattering resummation of gluon exchanges in an eikonal-type way representing the black disk limit of the proton. In the original model, the coupling constant is considered fixed as $\alpha_s=0.2$ and in order to consider the formal limit of photoproduction, the Bjorken variable has been modified in the following way,
\begin{eqnarray}
x=x_{Bj}\left( 1+ \frac{4m_q^2}{Q^2}\right)= \frac{Q^2+ 4m_q^2}{W^2}\,.\label{eq:17}
\end{eqnarray}

Starting from the $\rk_{\perp}$-factorization approach for the total photoabsortion cross section, but disregarding a transverse momentum dependence in the argument of the strong coupling constant and in the variable $x$, an analytical result for the Fourier transform between the spaces can be obtained,
\begin{eqnarray}
\sigma_{dip}(x,\rr)= \frac{2\pi}{3}\, \int \frac{d^2 \rk_{\perp}}{\rk_{\perp}^4} \, \alpha_s\, {\cal F}(x,\rk_{\perp}^2)\, \left(1- e^{i\,\rk_{\perp} \cdot\, \rr}\right)\, \left(1- e^{-i\,\rk_{\perp} \cdot \, \rr}\right)\,.\label{eq:18}
\end{eqnarray} 

Therefore, the Eq. (\ref{eq:18}) can be used to extract the unintegrated gluon function from the model in the $\rr$-space, once the dipole cross section has a finite limit at $\rr \rightarrow 0$, denoted as $\sigma^{(\infty)}_{dip}(x)$. It can be written as \cite{BGBK}
\begin{eqnarray}
\alpha_s\, {\cal F}(x,\rk_{\perp}^2) & = & \frac{3}{4\pi}\, \int \frac{d^2\rr}{(2\pi)^2}\, \exp \left(i\,\rk_{\perp}\cdot \, \rr \right)\,\left[\,  \sigma^{(\infty)}_{dip}(x) - \sigma_{dip}(x,\rr) \, \right]\,\rk_{\perp}^4 \,, \label{eq:19a}\\
& = &  \frac{3}{8\pi^2}\, \int_0^{\infty} dr \, r\, J_0(\rkn_{\perp}\,r)\, \left[\, \sigma^{(\infty)}_{dip}(x) - \sigma_{dip}(x,r) \,  \right]\,\rkn_{\perp}^4 \,. \label{eq:19b}
\end{eqnarray}

After writting down the expression for the unintegrated gluon function to be employed in the studies of the next section, some considerations are in order. The saturation model is a quite successful approach for the small-$x$ region and it was also extended to simultaneously describe the diffractive DIS. However, when $x \rightarrow 1$, the approach is no longer suitable and threshold correction factors should be introduced. The simplest way to implement this is to consider dimensional-cutting rules: for a subprocess having $n_{spec}$ spectators quarks which do not interact with the photon, the corresponding threshold factor is given by $(1-x)^{2n_{spec}-1}$
. For example, including light and charm quarks, the number of spectators is $n_{spec}=4$ and in our analysis we include the multiplicative correction factor $(1-x)^7$ in the unintegrated gluon function from the saturation model. In next section, such a correction provides a correct description of the fixed target energy region, whereas the results from the  original model remain unalterated at intermediate and high energies.

To our purposes in this investigation, we use the following parameters corresponding to the parameterization which includes the charm quark: $\sigma_0=29.12$ mb, $\lambda=0.277$ and $x_0=4.1\cdot 10^{-3}$ \cite{GBW}. From Eqs. (\ref{eq:19a}) and (\ref{eq:15}), the unintegrated gluon distribution from the original saturation model supplemented by the threshold factor is given by
\begin{eqnarray}
 {\cal F}(x,\rk_{\perp}^2) = \frac{3\,\sigma_0}{4\,\pi^2 \alpha_s} \, R_0^2(x)\, \rk_{\perp}^4 \exp \left( -R_0^2(x)\,\rk_{\perp}^2  \right) \, (1-x)^7\,.
\label{eq:21}
\end{eqnarray}

\begin{figure}[t]
\begin{center}
\begin{tabular}{cc}
\epsfig{file=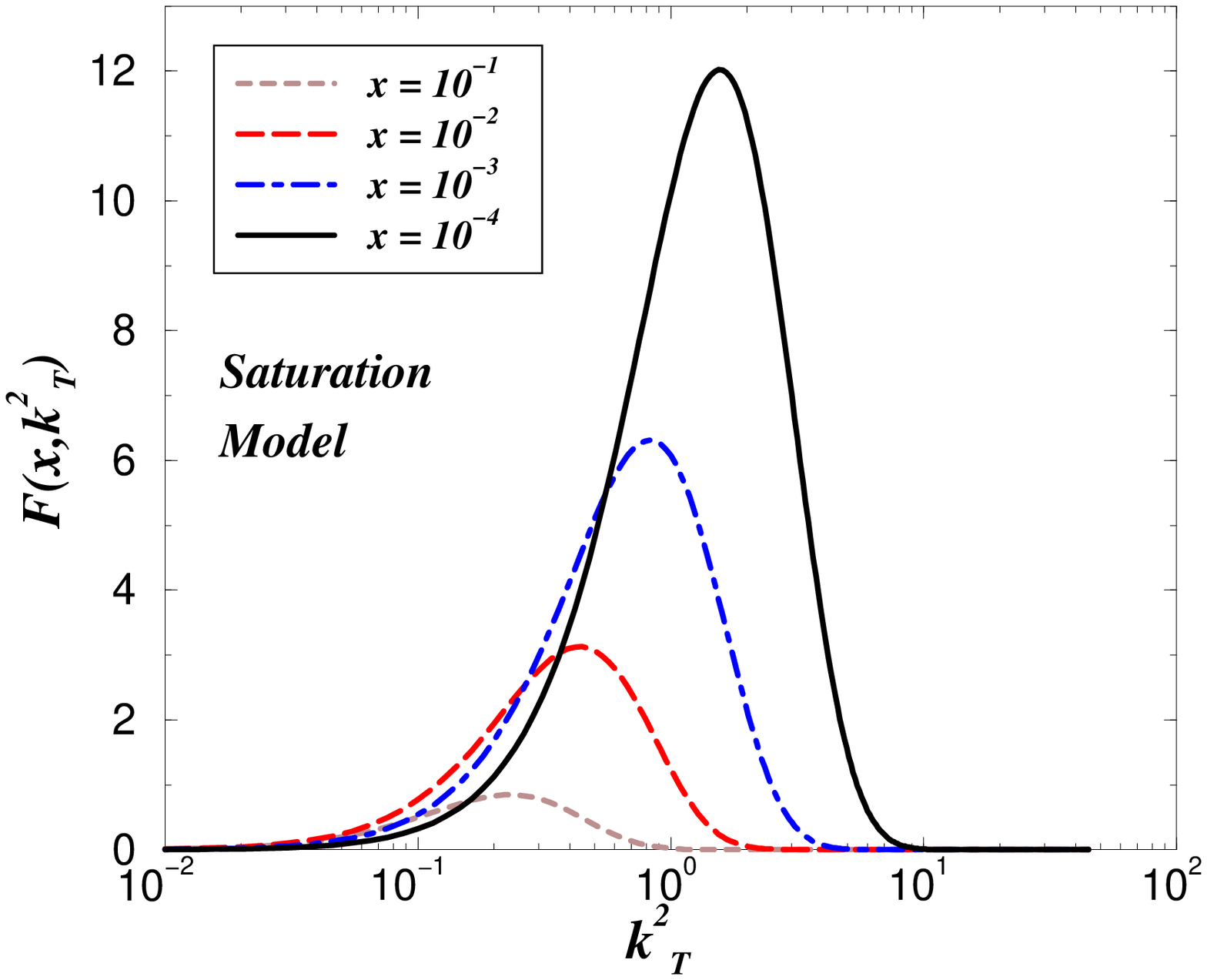,width=7.5cm} & 
\epsfig{file=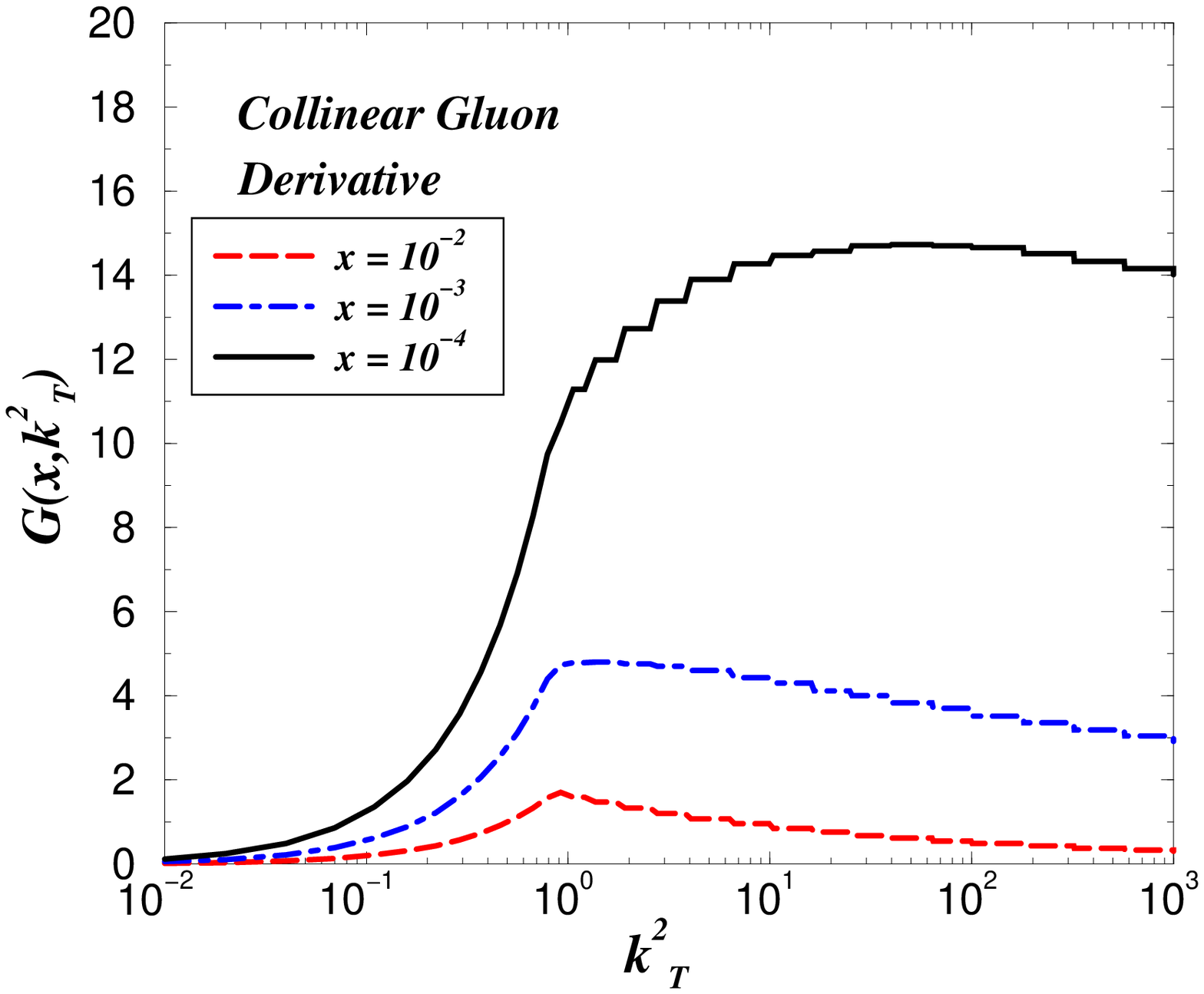,width=7.5cm}
\end{tabular}
\caption{\label{GBWunin} \underline{Left plot}: The unintegrated gluon distribution from the saturation model (GBW) as a function of $\rk_{\perp}^2$ for typical $x$ values. \underline{Right plot}: the derivative of the collinear gluon distribution (LO GRV98 parametrization)  as a function of $\rk_{\perp}^2$ for typical $x$ values. }
\end{center}
\end{figure}

We illustrate in Fig. (\ref{GBWunin}), shown in the plot on the left,  the unintegrated gluon function from Eq. (\ref{eq:21}) as a function of the transverse momentum $\rk_{\perp}^2$ at typical values of the $x$ variable, covering large and small longitudinal momentum fractions. We remark that the small-$x$ region corresponds to $x\leq 10^{-2}$. Its main features are quite clear: the function is peaked at $\rk_{\perp}^2=2\,Q_s^2$, with a narrow distribution around this value, being somewhat slightly asymmetric for large $x$. Therefore, the peak is shifted to larger $\rk_{\perp}^2$ as $x$ decreases. The most important feature is a large contribution from the very small transverse momentum  sector, $\rk_{\perp}\leq 1$ GeV$^2$ at large $x$. Moreover, the unintegrated gluon distribution is strongly suppressed for large $\rk_{\perp}^2$ due to the missing parton evolution in the original model. This shortcoming is cured in the recent implementation of DGLAP evolution for the unintegrated gluon distribution \cite{BGBK}.

In order to investigate the importance of a $\rk_{\perp}$-enhancement coming from a QCD parton cascade emission, we present in the plot on the right in Fig. (\ref{GBWunin}) the result from the derivative of the collinear gluon distribution, ${\cal G}(x,\rk_{\perp}^2)$, where it was chosen the LO GRV98 \cite{GRV98} gluon parametrization  for the calculation. The use of this quantity gives the possibility to
check the consistency of introducing elements from DGLAP evolution in the $\rk_{\perp}$-factorization approach. That parametrization has the virtuality $Q_0^2=0.8$ GeV$^2$ as initial evolution scale 
and bellow this value one should make an assumption for the behavior of ${\cal G}$.
In our analysis we apply the following procedure,
\begin{eqnarray}
{\cal G}(x,\rk_{\perp}^2) & = & \rk_{\perp}^2 \left. \frac{\partial\, [\,x\,G(x,\rk_{\perp}^2)\,]}{\partial \,\rk_{\perp}^2}\right |_{\rk_{\perp}^2=Q_0^2}\,\Theta ( Q_0^2 - \rk_{\perp}^2) \nonumber \\
& + &   \frac{\partial\, [\,x\,G(x,\rk_{\perp}^2)\,]}{\partial \, \ln \rk_{\perp}^2}\,\Theta (\rk_{\perp}^2 - Q_0^2)\,,\label{eq:22}
\end{eqnarray}
where the first term can recover $xG(x,Q_0^2)$ by integration over transverse momentum from 0 up to $Q_0^2$. However, we emphasize that such a procedure is not unique and other ansatze can be introduced. For instance, we can consider a lower integration limit $\rk_{\perp\, min}^2 \simeq Q_0^2$, emphasizing that in the theoretical curves the cut-off mainly would affect the overal magnitude of the cross section.  As expected, in our results the transverse momentum spectrum is broader, in contrast with the saturation model. At the small-$x$ region the deviation is huge both in behavior and in overall normalisation. The large $x$ region is correctly described since the collinear gluon function is adjusted on the whole kinematical range at HERA. It has been verified that by using the NLO collinear gluon distribution the deviations are negligible \cite{Jung:2001rp}. A technical remark is the pronged behavior of ${\cal G}(x,\rk_{\perp}^2)$ above $Q_0^2$, which has no physical meaning, since it has to do with the grid interpolation routines used to obtain the collinear gluon function, and this effect 
is smoothed out in the integrated quantities  and  should not affect our latter results.

In the next section we investigate the saturation model in the computation of the total and differential heavy-quark photoproduction cross section, and a comparison with the derivative of the collinear gluon function will be used 
in order to study the effects of QCD evolution.

\section{Results and discussions}
\label{sect:3}

The available data \cite{chafix, chahigh, botfix, bothigh} on heavy-quark (charm and bottom) photoproduction range from energies $W$ of fixed target experiments 
about tens of GeV, up to the high energy HERA data 
around $W\sim 200$ GeV. Kinematically, the low energy data corresponds to $x\simeq 10^{-1}$ and the high energy ones to $x\simeq 10^{-4}$. The experimental errors  are rather large and the intermediate region between low and high energies is not covered with measurements. Beauty production at HERA is suppressed by two orders of magnitude with respect to charm, 
due to the larger mass and smaller electric charge of the $b$ quark. 

The total and differential cross section are computed from Eq. (\ref{eq:11}) and the unintegrated gluon function from the saturation model, Eq. (\ref{eq:21}). The heavy quark mass 
was considered as $m_c=1.5$ GeV for charm and $m_b=4.8$ GeV for bottom. In order to investigate in detail the results emerging from the $\rk_{\perp}$-factorization approach and in an attempt to go beyond the leading logarithmic $\ln (1/x)$ approximation, we have considered in our analysis the following procedures:

\begin{figure}[t]
\begin{center}
\begin{tabular}{cc}
\epsfig{file=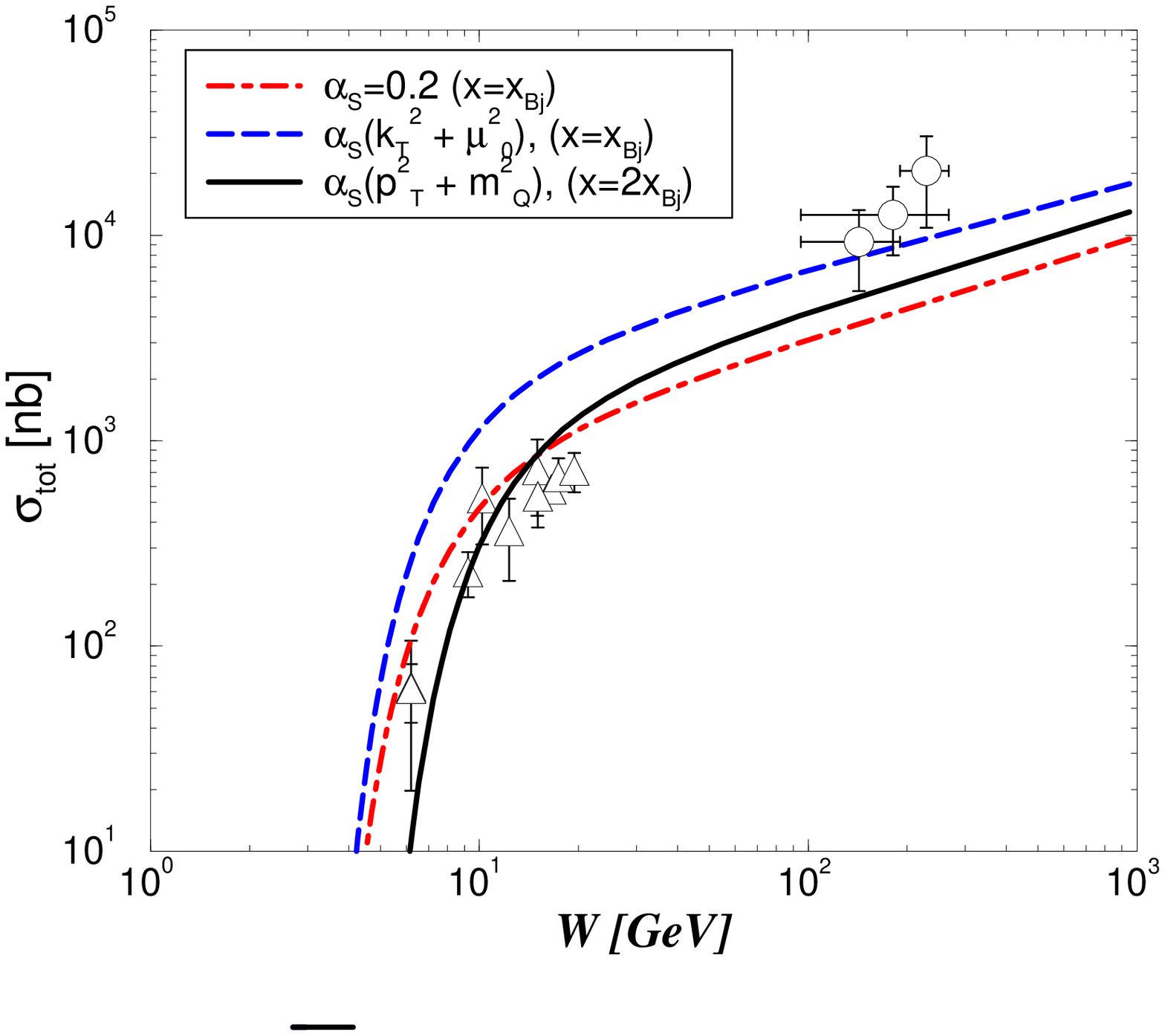,width=7.5cm} & 
\epsfig{file=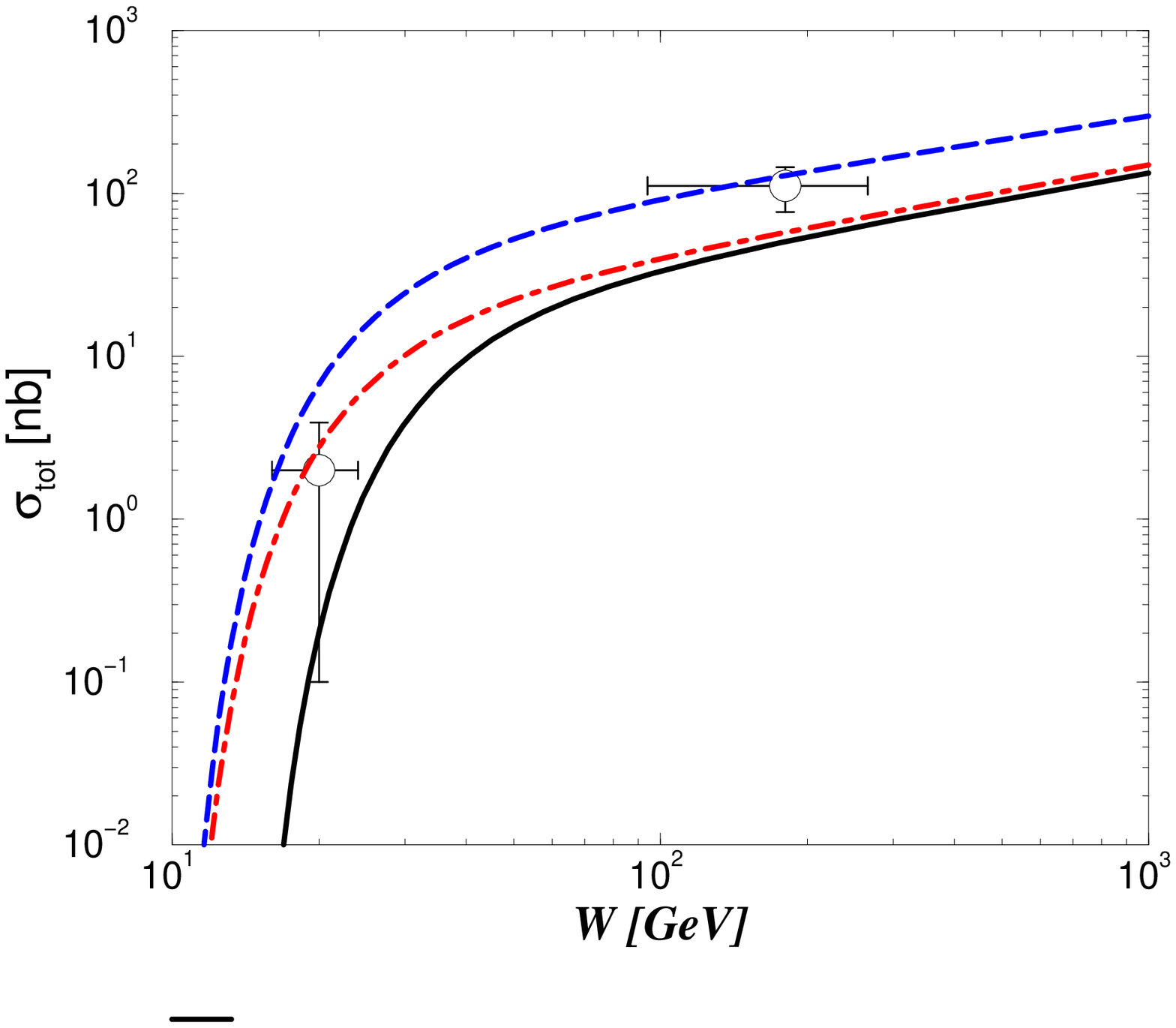,width=7.5cm}
\end{tabular}
\caption{\label{sigtots}  \underline{Left plot}: The charm photoproduction total cross section as a function of center of mass energy $W$ and the  $\rk_{\perp}$-factorization approach results using the saturation model and different technical procedures in the calculation (see text).  \underline{Right plot}:  The bottom  photoproduction total cross section as a function of center of mass energy $W$ and the  $\rk_{\perp}$-factorization approach results using the saturation model and different technical procedures in the calculation (see text).  }
\end{center}
\end{figure} 
\begin{enumerate}

\item We keep the original features of the saturation model. Namely, a fixed strong coupling constant $\alpha_s=0.2$ and the longitudinal momentum fraction $x_2$ being the photoproduction limit of the Bjorken $x_{Bj}$, given by Eq. (\ref{eq:17}), entering in the unintegrated gluon distribution. This procedure is equivalent to using 
the color dipole picture and the saturation model, 
 as performed in the detailed study of  Ref. \cite{Szczurek}. The result is shown in the dot-dashed curves in Fig. (\ref{sigtots}), for charm (plot on the left) and bottom (plot on the right) total cross sections. There is good agreement with the low energy data, whereas the high energy data are underestimated. The good description of the fixed target data is ensured by the threshold correction factor $(1-x)^7$, since the original model would overestimate the data already at intermediate energies. At high energies, those threshold effects are completely negligible at $W \gtrsim 20$ GeV for charm and $W \gtrsim 50$ GeV for bottom. Similar conclusions are obtained in the Ref. \cite{Szczurek}. One small difference between the procedure above and the calculations in  Ref. \cite{Szczurek} is the prescription for the longitudinal momentum fraction being $x_Q=m^2_Q/z(1-z)$ in the latter. If we consider a mean value $z=0.5$ for the quark momentum fraction, the results are completely equivalent. The very close similarity between our predictions and those of Ref. \cite{Szczurek} corroborates our procedure of calculation.

\item We allow the argument of the strong coupling constant to run with the following scale, $\mu^2=\rk_{\perp}^2 + \mu_0^2$. This procedure is still very close to the saturation model, and it is the general procedure in calculating observables by other groups \cite{Shabel-Shuva,LiSaZot2000}. The term $\mu_0^2$ is introduced in order to avoid divergencies coming from  the coupling constant in the infrared region. 
The value $\mu_0^2=1$ GeV$^2$ has been used, motivated by the value of the saturation scale $Q_s$, ensuring that the low transverse momentum region is dominated by that scale. The result seems to spoil the previous good agreement at low energies, as shown in the dashed curves in Fig. (\ref{sigtots}). However, the high energy data are described in better agreement than in the previous procedure. In conclusion, the introduction of a running coupling in the calculation shifts the overall normalisation towards higher values at high energies, by enhancing the $\rk_{\perp}^2$-profile in the unintegrated cross section.

\item As a final aspect, we consider a conservative procedure concerning the $\rk_{\perp}$-factorization approach. 
The  argument of the strong coupling constant is let to run with the scale $\mu^2=\rp_{\perp}^2 + m_Q^2$, where $\rp_{\perp}$ is the transverse momentum of the produced heavy quark. Moreover, we have a different prescription for the $x_2$ variable than in Eq. (\ref{eq:17}). The correct value for the longitudinal momentum fraction entering in ${\cal F}(x_2,\rk_{\perp}^2)$ is given by the definitions in Eqs. (\ref{eq:8}) and (\ref{eq:7}). Namely, the momentum fraction is given by
\begin{eqnarray}
x_2 = \frac{m_{1\perp}^2}{z\,W^2} + \frac{m_{2\perp}^2}{(1-z)\,W^2}= \frac{m_Q^2+(\rp_{\perp}-\rk_{\perp})^2}{z}+ \frac{m_Q^2+\rp_{\perp}^2}{1-z} \,, \label{eq:23}
\end{eqnarray}
where one has made use of the relations in Eqs. (\ref{eq:6}) and (\ref{eq:7}) and the energy-momentum conservation law, Eq. (\ref{eq:5}). Although Eq. (\ref{eq:23}) is well defined, it is involved for practical computations and for simplicity we rely on the following approximation. 
We benefit from the results on the $\rk_{\perp}$-factorization approach applied to $ep$ collisions, in particular for the proton structure function, performed in Ref. \cite{IvaNiko}. There, it was verified that for not too high virtualities $Q^2$ (including photoproduction), a suitable approximation is $x_2=2\,x_{Bj}$. This value is obtained from a careful investigation of the contributions in the transverse momentum integration range for the corresponding DGLAP piece and for the  semihard approach. The results using the procedure above is shown in the solid lines in Fig. (\ref{sigtots}), presenting an intermediate behavior between the first and the second procedures. In particular, the fixed target data are described in good agreement and the results for high energy data are slightly improved compared to the equivalent dipole result (specially for the charm results). 
\end{enumerate}

A more detailed calculation considering the resolved part of the photon is beyond the scope of
the present analysis. In Ref. \cite{Shabel-Shuva},  
the resolved component is considered by including 
the off-shell matrix elements identical to those of heavy
quark hadroproduction, convoluted with the photon and the proton unintegrated parton densities. 
In Ref. \cite{Szczurek}, where the color dipole picture is applied to heavy-quark photoproduction using the saturation model, a VDM contribution is included. In our analysis, we only consider the matrix elements of the direct component of the photon. According to the authors of Ref. \cite{BaranovJJPZ}, the $\gamma \to c\bar{c}$ component of the photon is automatically included in the $\rk_{\perp}$-factorization approach,
since there is no restriction on the transverse momenta along the evolution chain. 
Moreover, it is shown in Ref. \cite{BaranovJJPZ}, in particular when calculating  the $D^*$ (mesons) differential cross section $d\sigma/dx_{\gamma}$, that part of the resolved photon contribution is effectively included by the BFKL or CCFM \cite{CCFM} evolutions, namely this is included in the evolution of the unintegrated gluon distribution, and the off-shell matrix elements would contain only the direct component of the photon. 
When using such approaches in photoproduction, the resolved photon contribution is in general not included explicitly in order to avoid double counting, since the off-shell gluon from the BFKL evolution would take already into account a certain  portion of this contribution. 

In order to investigate the influence of the procedures above in the predictions for the total cross section and to find from where (in the transverse momentum range) comes the main contribution, we propose to consider the $\rk_{\perp}$ profile of the unintegrated cross section. This quantity, denoted by $W(x,\rk_{\perp}^2)$, is obtained by unfolding the $\rk_{\perp}^2$-integration in Eq. (\ref{eq:11}).
\begin{figure}[t]
\begin{center}
\begin{tabular}{cc}
\epsfig{file=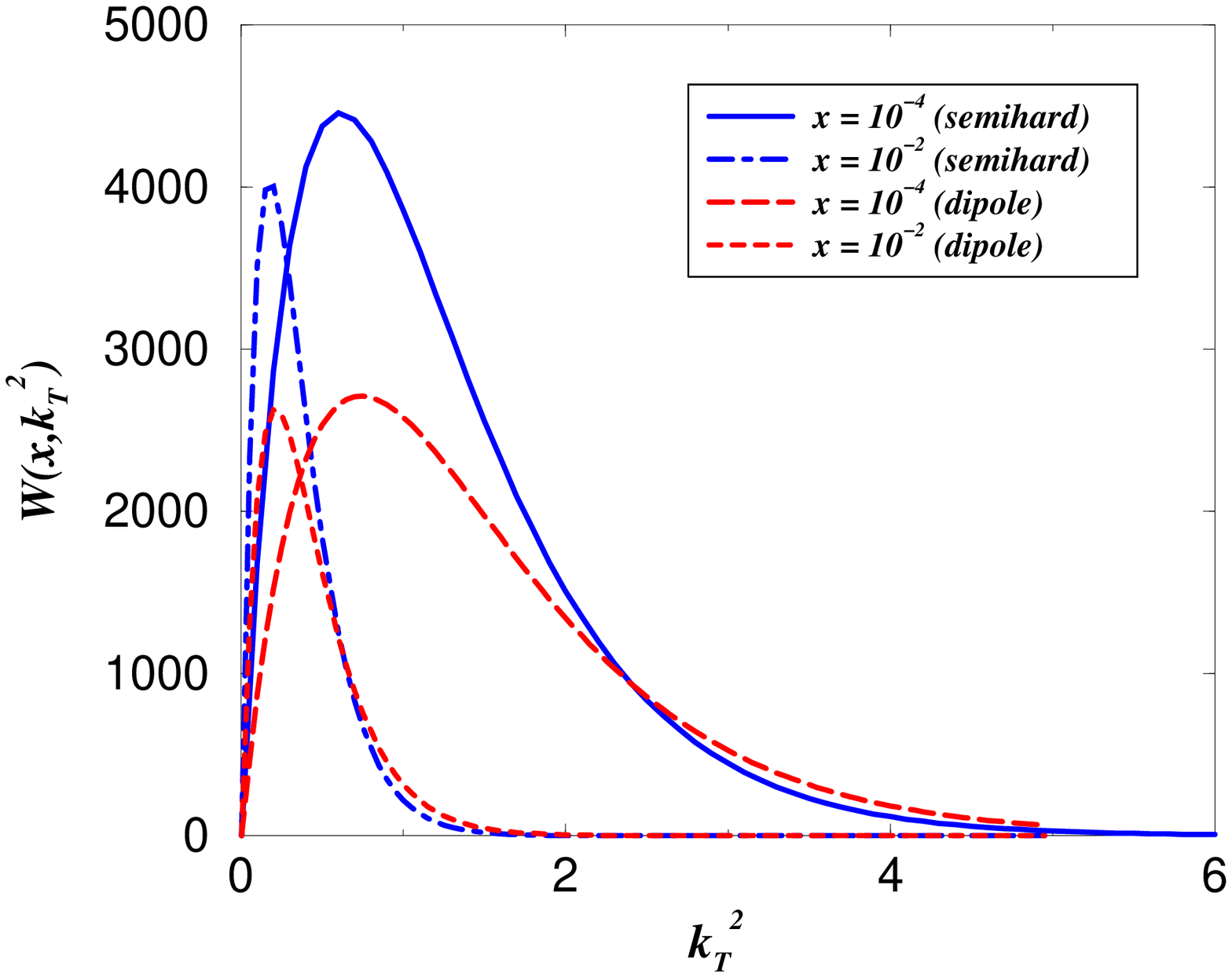,width=7.5cm} & 
\epsfig{file=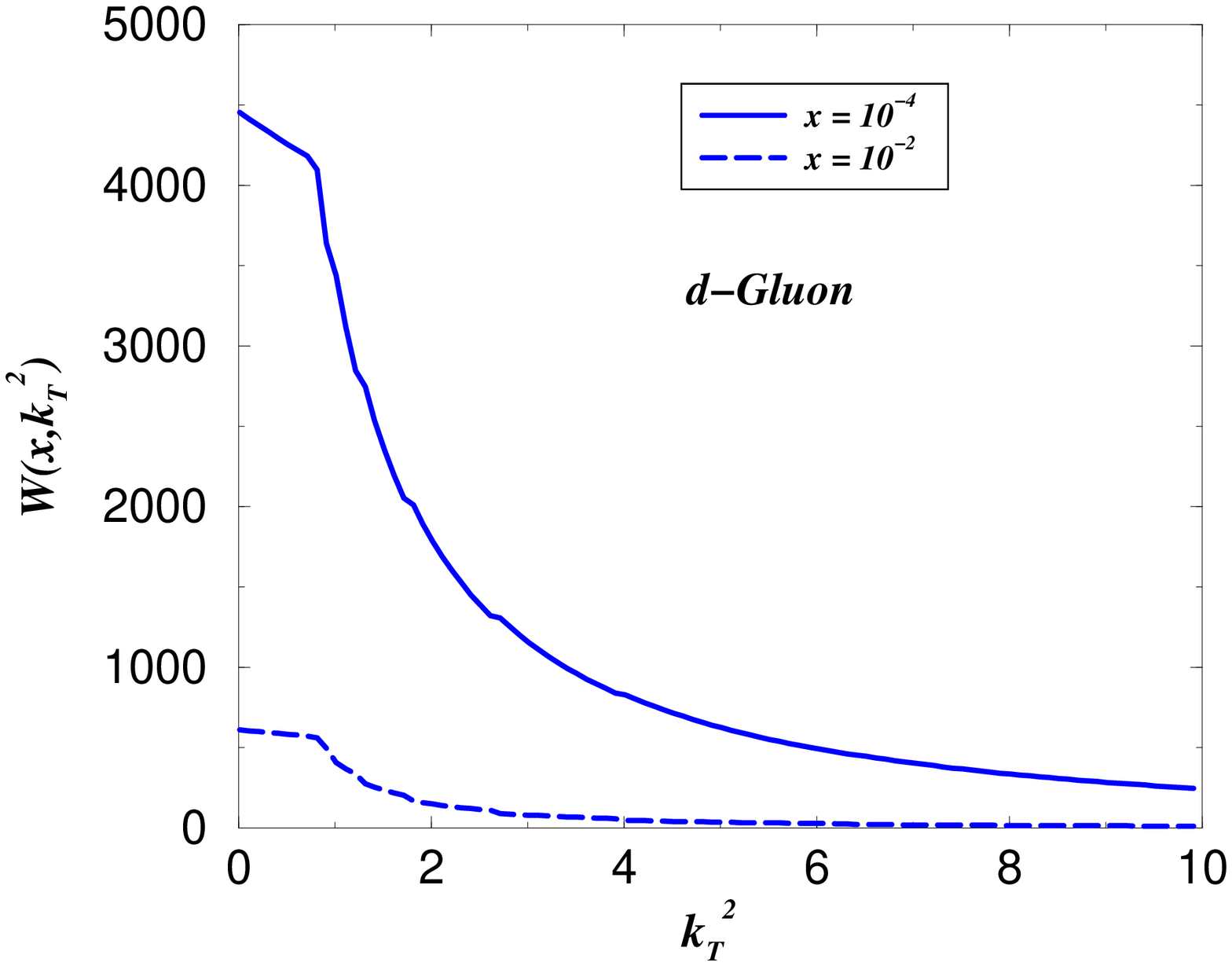,width=7.5cm}
\end{tabular}
\caption{\label{profiles} \underline{Left plot}: The $\rk_{\perp}$ profile (charm) for the unintegrated cross section from the saturation model for two different  procedures concerning coupling constant and momentum fraction scale (see text). \underline{Right plot}:  The $\rk_{\perp}$ profile (charm) for the unintegrated cross section from the derivative of the collinear gluon distribution (LO GRV98 parametrization). }
\end{center}
\end{figure}    
In Fig. (\ref{profiles}) we show the profile functions for charm, by using the saturation model (plot on the left) and the derivative of the collinear gluon distribution (plot on the right), for fixed target energies (momentum fraction $x=10^{-2}$) and high energies ($x=10^{-4}$). 
For the saturation model, we consider two of the procedures above, namely the dipole approximation (fixed $\alpha_s$ and $x_2=x_{Bj}$) and the semihard approach ($\alpha_s$ running with $\mu^2=\rp_{\perp}^2+m_Q^2$ and $x_2=2x_{Bj}$). 
At $x=10^{-2}$ the profile functions are peaked around $\rk_{\perp}^2 \simeq 0.3$ GeV$^2$, whereas at $x=10^{-4}$ the peaks are shifted towards $\rk_{\perp}^2 \simeq 1$ GeV$^2$.
As expected from results for the total cross section (Fig. \ref{sigtots}), the semihard approach results have a larger normalization than the dipole ones. An important feature emerging from these results is the dominance of the small $\rk_{\perp}^2$ region for the charm total cross section. Indeed, at high energies the peak is of order of the saturation scale, $\rk_{\perp}^2 \simeq Q_s^2$, and contributions for $\rk_{\perp}^2 \gtrsim 10$ GeV$^2$ are negligible.

For the derivative of the collinear gluon distribution (usually denoted $d-Gluon$), we choose the scale $\mu^2=\rp_{\perp}^2 + m_Q^2$ and $x_2=2\,x_{Bj}$. We can notice in Fig. (\ref{profiles}b) the effect of the discontinuity in the derivative of the gluon function, Eq. (\ref{eq:22}) at $Q_0^2=0.8$ GeV$^2$: the profile function peaks in this value for $\rk_{\perp}^2 \geq Q_0^2$, whereas it has a flatter behaviour for $\rk_{\perp}^2$ below $Q_0^2$.
We can also notice that the $\rk_{\perp}$ profile for the derivative of gluon distribution is broader than in the saturation model. Still the main contribution comes from the small $\rk_{\perp}$ region, however, intermediate values of transverse momentum give a considerable contribution. 
This will bring the predictions closer to data, as we will see latter on.
\begin{figure}[t]
\begin{center}
\begin{tabular}{cc}
\epsfig{file=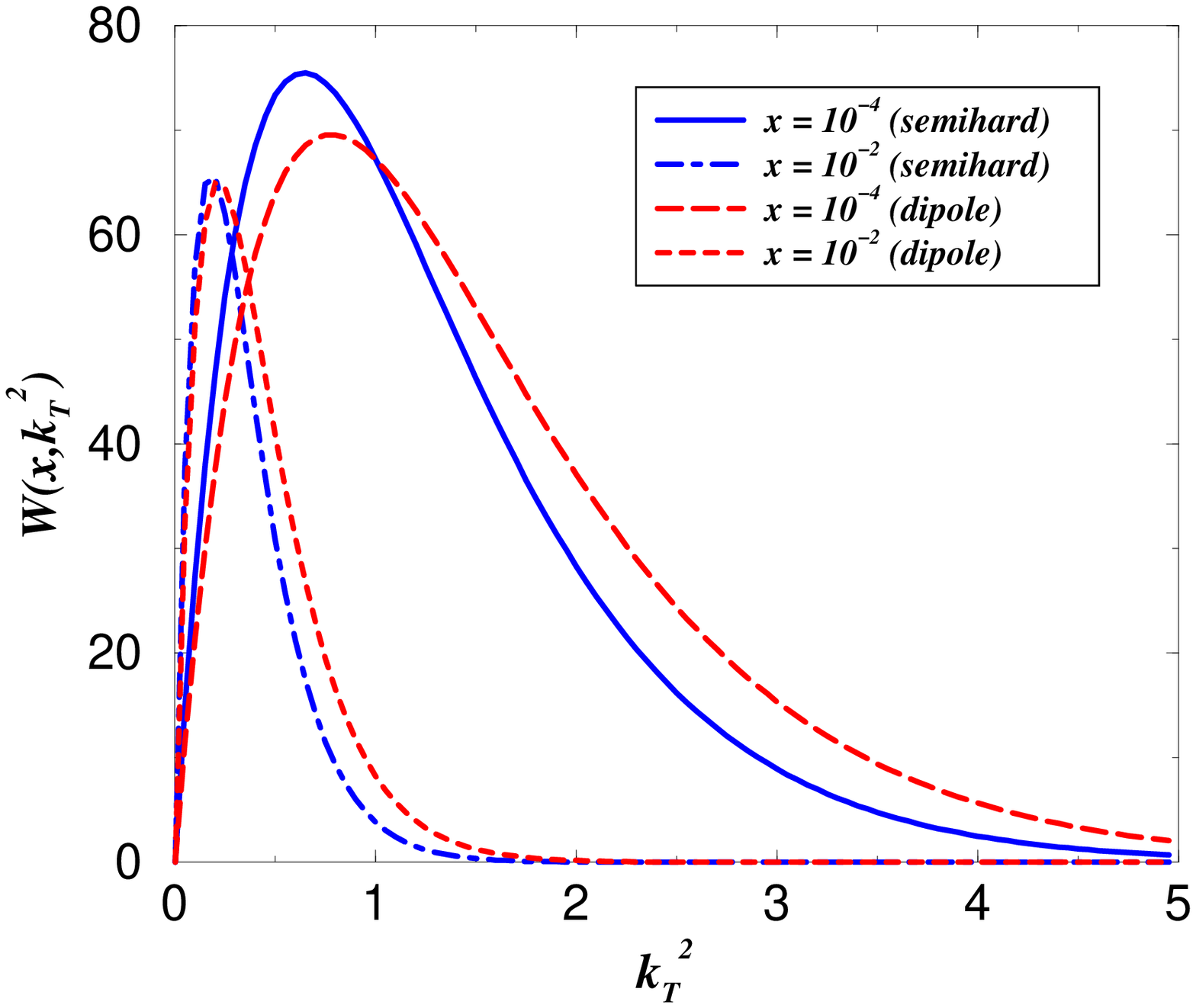,width=7.5cm} & 
\epsfig{file=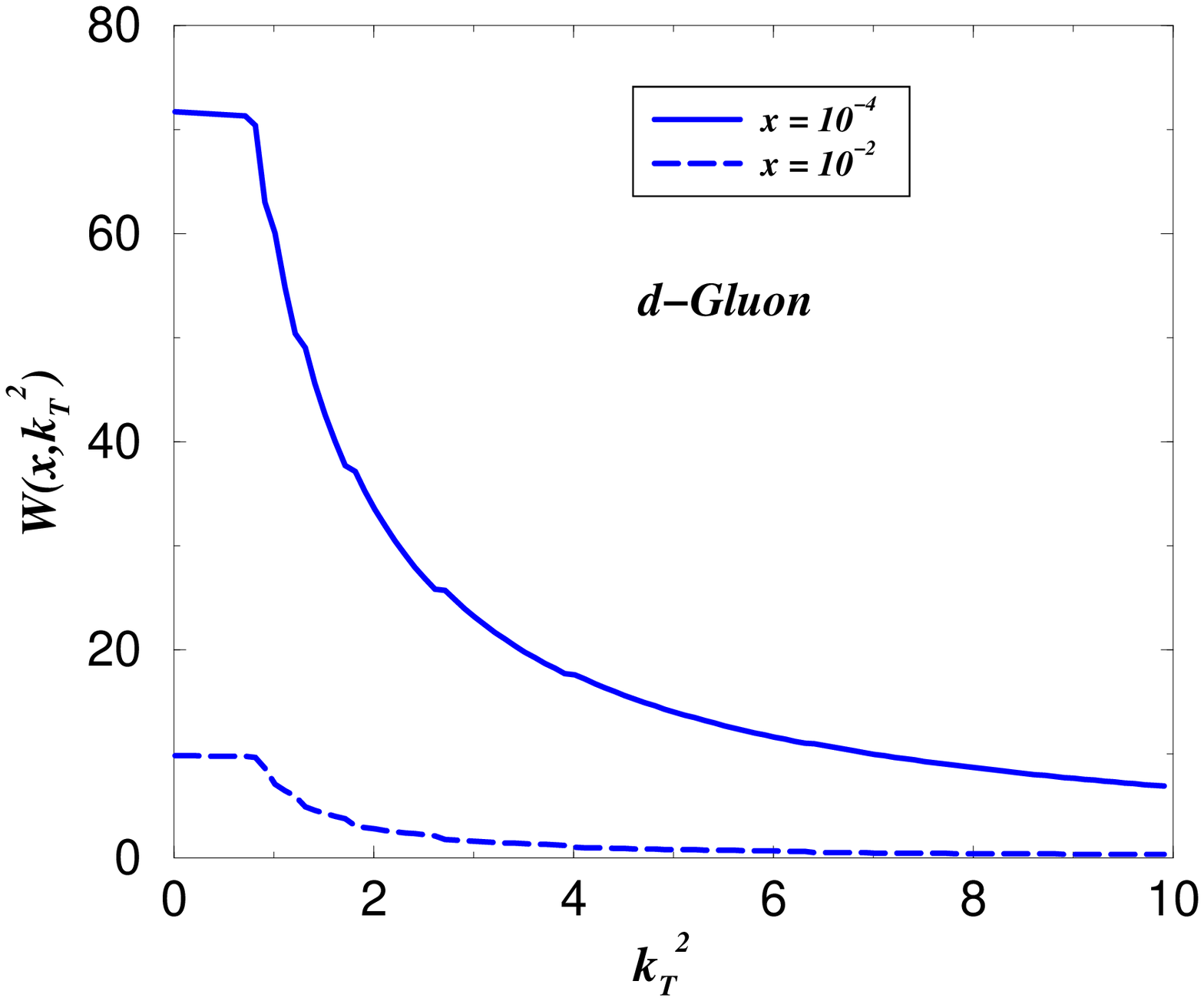,width=7.5cm}
\end{tabular}
\caption{\label{profilesbottom} \underline{Left plot}: The $\rk_{\perp}$ profile (bottom) for the unintegrated cross section from the saturation model for two different  procedures concerning coupling constant and momentum fraction scale (see text). \underline{Right plot}:  The $\rk_{\perp}$ profile (bottom) for the unintegrated cross section from the derivative of the collinear gluon distribution (LO GRV98 parametrization). }
\end{center}
\end{figure}    

In Fig. (\ref{profilesbottom}) we present the results for the $\rk_{\perp}$ profile for the bottom, where we compare the different procedures and unintegrated gluon functions, similarly to the charm case. Some differences in comparison with charm are evident. For the saturation model calculations, the results of the dipole and semihard procedure are very similar, since the scale $\mu^2=\rp_{\perp}^2+ m_Q^2$ provides a high virtuality even at very small transverse momentum due to the large mass of the bottom. 
This makes the strong coupling constant to be very close to the value $\alpha_s=0.2$ in the whole $\rp_{\perp}$ range, as assumed in the original saturation model. 
The results from the derivative of the collinear gluon distribution keep the same features as the charm case, having again a broader $\rk_{\perp}$ profile in comparison with the saturation model.    

\begin{figure}[t]
\centerline{\psfig{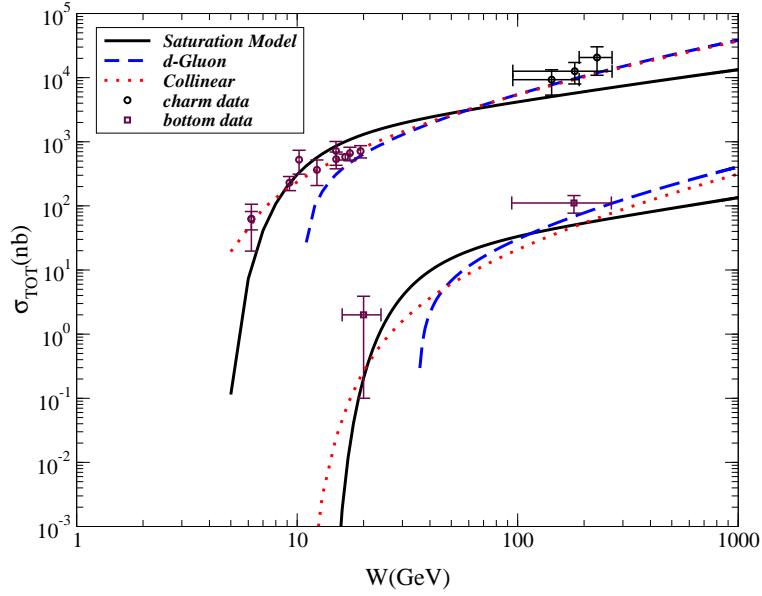}}
\caption{The results for the charm and bottom total cross sections considering the saturation model, the derivative of the collinear gluon distribution and the collinear parton model. See discussion in the text.}
\label{fig:sigtot} 
\end{figure}

The study of the $\rk_{\perp}$ profile 
shows the dominant region in the $\rk_{\perp}$ range and the effect of choosing distinct scales for the coupling and longitudinal momentum fractions. From the profiles discussed above, we expect that the broader $\rk_{\perp}$ spectrum for the derivative  of the collinear gluon function will enhance the overall normalisation of the total cross section at high energies, improving the data description in comparison with the saturation model predictions. Motivated by this fact, we also perform a comparison at total cross section level between the two unintegrated gluon functions, namely the saturation approach and the derivative of the collinear gluon funtion. In both cases we make our default choice of scale $\mu^2=\rp_{\perp}^2+ m_Q^2$ and $x_2=2\,x_{Bj}$. 
This comparison is shown in Fig. (\ref{fig:sigtot}) for both charm and bottom total cross sections. 
The saturation model underestimates high energy data, since the treatment of QCD evolution is not considered in the original model. Recent improvements, taking QCD evolution into account, should cure this shortcoming \cite{BGBK}. The derivative of the collinear gluon distribution gives a better description of high energy data, since it includes the referred gluon emission. As expected, it is in disagreement at lower energies, since the non-singlet (valence) content was not included in the analysis. In addition, it was also verified that the corresponding unintegrated distribution function takes negative values in that region. For sake of illustration, we also show the parton model results (collinear approach) for the LO process $\gamma g \rightarrow Q\bar{Q}$, where it has been used $m_c=1.3$ GeV, $m_b=4.75$ GeV, and $\mu^2=\hat{s}$.
This gives a reasonable description of data given the use of lower heavy quark masses or alternativelly considering higher order corrections to the LO calculation.
In contrast, the semihard approach gives a reasonable 
description of data already at LO level. 
The energy dependence is distinct in the calculations: the saturation model provides a mild energy growth, whereas in the collinear approach the growth is steeper. The collinear approach and the semihard result using the derivative of gluon function present a similar energy behavior, with sizeable deviations only at low energies near the threshold. 

\begin{figure}[t]
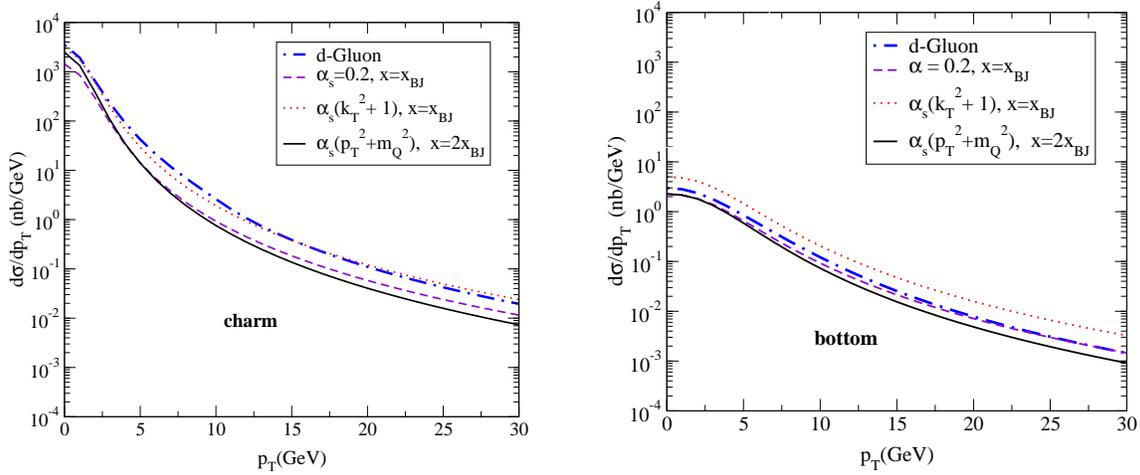

\begin{center}
\begin{tabular}{cc}
\epsfig{file=dsigdptcharpap.eps,width=75mm} &
\epsfig{file=dsigdptbotpap.eps,width=82mm}
\end{tabular}
\caption{\label{ptdistrib} \underline{Left plot}: The charm $\rp_{\perp}$-distribution from the saturation model (using three distinct procedures) and  the derivative of the collinear gluon distribution for $W=200$ GeV. \underline{Right plot}:  The bottom $\rp_{\perp}$-distribution from the saturation model (using 3 distinct procedures) and  the derivative of the collinear gluon distribution for $W=200$ GeV. }
\end{center}
\end{figure}

As a final investigation, we compute the $\rp_{\perp}$-distribution for charm and bottom using the saturation model (using the 3 procedures discussed earlier) and the derivative of the collinear gluon distribution, at center of mass energy $W=200$ GeV. The predictions are shown in the Fig. (\ref{ptdistrib}). A remarkable feature is
the finite and well controlled behavior at small $\rp_{\perp}$ for both gluon functions. One can also see the usual fall off at large transverse momentum.
The finiteness at zero transverse momentum in a LO level calculation is one of the main advantages of the semihard approach. Our results are comparable with those of Refs.\cite{Shabel-Shuva,LiSaZot2000}, which consider other parametrizations for the unintegrated gluon function. Our results for the saturation model are quite similar, even using different prescriptions for the scales at $\alpha_s$ and for longitudinal momentum fraction, with a slightly deviation at larger transverse momentum. The growth at small $\rp_{\perp}$ is less steep in the bottom case than in the charm calculation, because of the larger bottom mass in the argument of $\alpha_s$. The derivative of the collinear gluon distribution ($d-Gluon$) produces a similar behavior on $\rp_{\perp}$, but with a somewhat higher overall normalization.

\section{Conclusions}

In this paper we investigate in detail the $\rk_{\perp}$-factorization approach (semihard approach) applied to 
heavy-quark photoproduction. In this formalism the cross section is given by the convolution of off-shell matrix elements with the unintegrated parton distributions, ${\cal F}(x_2,\rk_{\perp}^2)$. The matrix elements are at present known at LO accuracy, and include most of the NLO and even NNLO diagrams from the collinear factorization approach. This fact is advantageous in heavy-quark production, since 
NLO calculations in the collinear approach undershoot data (specially Tevatron data for bottom).
There are several parameterizations for the unintegrated gluon function, relying on the solution of evolution equations or based on phenomenological considerations.
We have investigated the application of the 
saturation model parameterization, which provide us a safe treatment of the infrared region and includes the onset of the parton saturation phenomenon.  Moreover, the adjustable parameters of this model are extracted from the high energy HERA data, and therefore the results are parameter free.

In order to go beyond the leading logarithmic approximation ($\ln (1/x)$), we let 
the strong coupling constant run and use a suitable longitudinal momentum fraction entering on ${\cal F}$. The description of the total cross section data is strongly dependent on those procedures. An additional ingredient in the calculations is a threshold correction factor accounting for the low energy behavior. The best agreement is obtained by the prescription $\mu^2=\rp_{\perp}^2+ m_Q^2$ and $x_2=2\,x_{Bj}$. However, in any case the saturation model slightly underestimate the high energy experimental results. This comes from the fact that QCD evolution is not present in the original model. In order to investigate the role played by these evolution emissions, we have considered the derivative of the collinear gluon distribution, which provides a closer connection with the DGLAP formalism. Indeed, the results for the charm and bottom total cross sections are in nice agreement in both low and high energies. Recent improvements of the saturation model considering  these emissions are expected to produce similar results.

It was verified that the study of the $\rk_{\perp}$ profile provides important information where in the range of transverse momentum the main contribution for the processes is coming from. For the saturation model the most important piece is peaked at the saturation scale $\rk_{\perp}^2\simeq Q_s^2$. The $\rk_{\perp}$ spectrum is broader for the derivative of the gluon distribution, ${\cal G}(x_2,\rk_{\perp}^2)$.
The $\rp_{\perp}$-distribution of the produced heavy-quarks is also computed, showing the effects coming from different prescriptions for the scales considered. The results for the saturation model are quite similar, even using different prescriptions for the scales at $\alpha_s$ and for the longitudinal momentum fraction, with a slight deviation at large transverse momentum. The growth at small $\rp_{\perp}$ is less steep in the botton case than for the charm. The derivative of the collinear gluon distribution produces a similar behavior on $\rp_{\perp}$.

The study of heavy quark photoproduction in the framework of the semihard approach improves the understanding of QCD dynamics both in the infrared and in the perturbative regions. It also sheds light on the proton structure, specially the gluon distribution. Therefore, it contributes in the understanding of the amazing interplay of soft and hard QCD phenomena.

\ack
This work was supported by CNPq, Brazil. MVTM is grateful to Prof. Nikolai Zotov (Skobeltsyn I.N.P., Moscow State University) and Antoni Szczurek (Institute of Nuclear Physics, Cracow) for useful enlightenements.

\end{document}